\documentclass[aip,amsmath,amssymb,reprint]{revtex4-1}

\usepackage{SIunits}
\usepackage{hyphenat}
\usepackage[bookmarks=false]{hyperref}
\usepackage{color}
\usepackage[capitalise]{cleveref}
\crefname{section}{Sec.}{Secs.}
\Crefname{section}{Section}{Sections}
\usepackage{floatrow}
\usepackage{graphicx}
\usepackage[caption=false]{subfig}
\definecolor{pink}{RGB}{255,0,255}

\definecolor{red}{rgb}{0,0,1}

\begin{document} 

\title{Optical control of single-photon negative-feedback avalanche diode detector}

\author{Ga\"etan~Gras}
\email{gaetan.gras@idquantique.com}
\affiliation{ID Quantique SA, CH-1227 Carouge, Switzerland}
\affiliation{Group of Applied Physics, University of Geneva, CH-1211 Geneva, Switzerland}

\author{Nigar~Sultana}
\email{n6sultan@uwaterloo.ca}
\affiliation{Institute for Quantum Computing, University of Waterloo, Waterloo, ON, N2L~3G1 Canada}
\affiliation{\mbox{Department of Electrical and Computer Engineering, University of Waterloo, Waterloo, ON, N2L~3G1 Canada}}

\author{Anqi~Huang}
\email{angelhuang.hn@gmail.com}
\affiliation{Institute for Quantum Information \& State Key Laboratory of High Performance Computing, College of Computer, National University of Defense Technology, Changsha 410073, People's Republic of China}
\affiliation{Institute for Quantum Computing, University of Waterloo, Waterloo, ON, N2L~3G1 Canada}
\affiliation{\mbox{Department of Electrical and Computer Engineering, University of Waterloo, Waterloo, ON, N2L~3G1 Canada}}

\author{Thomas~Jennewein}
\affiliation{Institute for Quantum Computing, University of Waterloo, Waterloo, ON, N2L~3G1 Canada}
\affiliation{Department of Physics and Astronomy, University of Waterloo, Waterloo, ON, N2L~3G1 Canada}

\author{F\'elix~Bussi\`eres}
\affiliation{ID Quantique SA, CH-1227 Carouge, Switzerland}
\affiliation{Group of Applied Physics, University of Geneva, CH-1211 Geneva, Switzerland}

\author{Vadim~Makarov}
\affiliation{Russian Quantum Center, Skolkovo, Moscow 121205, Russia}
\affiliation{\mbox{Shanghai Branch, National Laboratory for Physical Sciences at Microscale and CAS Center for Excellence in} \mbox{Quantum Information, University of Science and Technology of China, Shanghai 201315, People's Republic of China}}
\affiliation{NTI Center for Quantum Communications, National University of Science and Technology MISiS, Moscow 119049, Russia}
\affiliation{Department of Physics and Astronomy, University of Waterloo, Waterloo, ON, N2L~3G1 Canada}

\author{Hugo~Zbinden}
\affiliation{Group of Applied Physics, University of Geneva, CH-1211 Geneva, Switzerland}

\date{\today} 

\begin{abstract}
We experimentally demonstrate optical control of negative-feedback avalanche diode (NFAD) detectors using bright light. We deterministically generate fake single-photon detections with a better timing precision than normal operation. This could potentially open a security loophole in quantum cryptography systems. We then show how monitoring the photocurrent through the avalanche photodiode can be used to reveal the detector is being blinded.
\end{abstract}

\maketitle

\section{Introduction}

Quantum key distribution (QKD) allows two parties, Alice and Bob, to share a secret key. The first proposal of QKD was done by Bennett and Brassard in 1983 \cite{bennett1984}. Since then, this field has evolved rapidly. Unlike classical cryptography that makes assumptions on the computational power of an eavesdropper Eve, security proofs in QKD are based on the laws of quantum mechanics \cite{lo1999,shor2000}. 

However, imperfections in practical systems can open loopholes that can be used by a malicious third party to get some information on the key. Attacks of various types have been proposed, for example photon number splitting (PNS) attack \cite{huttner1995}, detector efficiency mismatch attack \cite{makarov2006}, Trojan-horse attack \cite{jain2014}, time-shift attack \cite{zhao2008}. In this paper, we are interested in a detector blinding attack, which belongs to the class of faked-state attacks \cite{makarov2005}. In this attack, Eve uses bright light to take control of the detectors in the QKD system to force the outcome of the measurement to be the same has her own. Such blinding on individual detectors has been demonstrated for single-photon avalanche diodes (SPADs) \cite{sauge2011,makarov2009,lydersen2010a,lydersen2011,gerhardt2011} and for superconducting nanowire single-photon detectors (SNSPDs) \cite{fujiwara2013,tanner2014,lydersen2011c}. 

Here, we show that negative-feedback avalanche diode (NFAD) detectors can be controlled with bright light. Such detectors are promising thanks to their high efficiency and low afterpulsing probability \cite{korzh2014}. We also show that diode current monitoring can be used to uncover the presence of blinding. We have tested 4 diodes made by Princeton Lightwave \cite{itzler2010}. Two of them are integrated in a commercial single-photon detector from ID~Quantique (model ID220 \cite{idq-id220}) and two are used with a custom readout circuit made at the University of Waterloo \cite{sultana2019}. 

\section{Experimental setup}

\begin{table}
\caption{Characteristics of our NFAD devices \cite{itzler2010}.}
\begin{tabular*}{\columnwidth}[t]{@{\extracolsep{\fill}}cccc}
   \hline
   \hline
   Designation & Model number & Diameter ($\micro\meter$) & Coupling\\
   \hline
   D1 & E2G6 & 22 & Capacitive\\
   D2 & E3G3 & 32 & Capacitive\\
   D3 & E2G6 & 22 & Inductive\\
   D4 & E3G3 & 32 & Inductive\\
   \hline
   \hline
\end{tabular*}
\label{tab:diodes}
\end{table} 

\begin{figure}
	\includegraphics{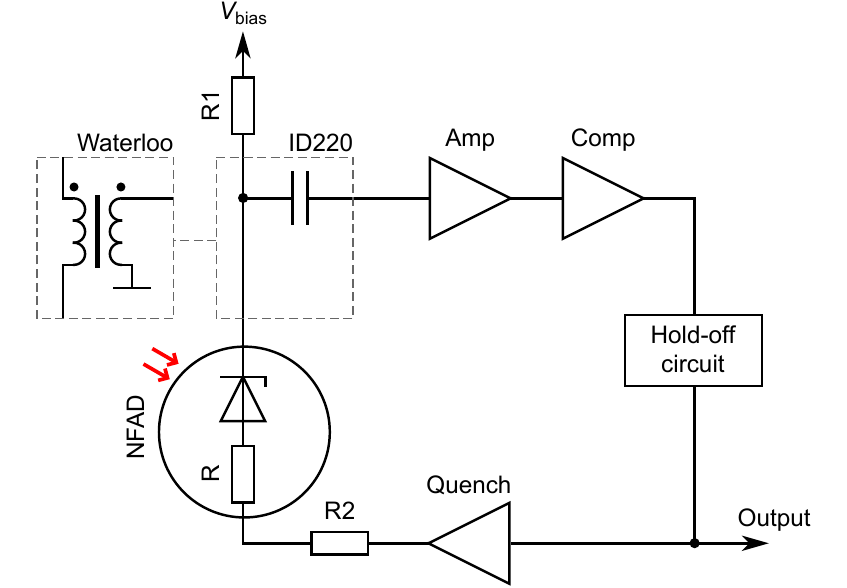}
	\caption{Scheme of the electrical readout. After detection of a photon by the APD, the avalanche signal is coupled to an amplifier (Amp) through a capacitor in ID220 or a pulse transformer in custom readout (Waterloo). Then it goes through a comparator (Comp). The hold-off circuit outputs a gate with a pre-set width. The feedback loop is used to quench the avalanche by applying a +5~V (ID220) or +4 V (custom readout) voltage to the anode of the NFAD for deadtime $\tau_d$. By applying this voltage, we reduce the voltage across the APD below its breakdown voltage. R $=1.1~\mega\ohm$ is a resistor integrated into the NFAD \cite{itzler2010}. In ID220, R1 $=1~\kilo\ohm$ and R2 $=50~\ohm$; for Waterloo, R1 $=1~\kilo\ohm$ and R2 $=100~\ohm$.}
	\label{fig:NFADcircuit}
\end{figure}

The characteristics of the four NFAD devices are given in \cref{tab:diodes}. The electronic circuit of the detectors is shown in \cref{fig:NFADcircuit}. It is similar for both setups except for the coupling to the amplifier, which is capacitive in D1 and D2, and inductive in D3 and D4. This differing part of the circuit is shown in dashed boxes. Under normal conditions, the NFAD works in Geiger mode, i.e.,\ the avalanche photodiode (APD) is biased with a voltage $V_\text{bias}$ greater than the breakdown voltage $V_\text{br}$. When a photon is absorbed, it creates an avalanche generating an electrical pulse. This analog signal is then converted into a digital signal by using a comparator with a threshold voltage $V_\text{th}$. To take control of the detector, Eve needs first to blind it so it becomes insensitive to single photons \cite{lydersen2010a}. To do so, she sends continuous bright light on the APD, which then generates a photocurrent. As the APD is connected in series with resistors R, R1, and R2 (see \cref{fig:NFADcircuit}), the voltage across the APD will be reduced. If Eve sends enough light she can then bring the voltage across the APD below $V_\text{br}$ and put the detector into a linear mode. In this mode, the detector is no longer sensitive to single photons but instead works as a linear photodetector. Eve can now force the detector to click at the time of her choosing by superimposing optical pulses (trigger pulses) to her blinding laser.

\begin{figure}
	\includegraphics{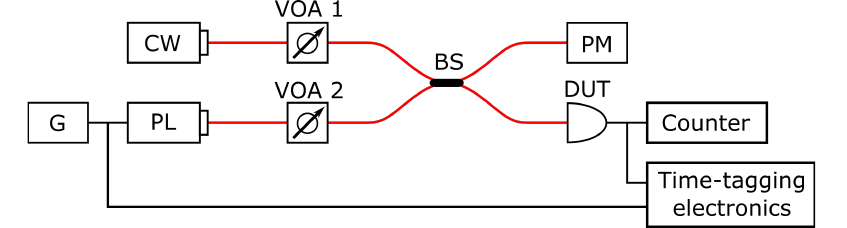}
	\caption{Experimental setup for testing blinding and control of the detectors. The optical power of the continuous-wave laser (CW) and the pulsed laser (PL) is adjusted using variable optical attenuators (VOA). The pulsed laser is triggered by a pulse generator (G). The two lasers are combined on a $50\!:\!50$ beam splitter (BS). The light is sent to the device-under-test (DUT) and to a power meter (PM).}
	\label{fig:expsetup}
\end{figure}

To test for blinding and control, we use a setup shown in \cref{fig:expsetup}. For the attack, we use two lasers at $1550~\nano\meter$ \cite{lydersen2010a}. The first laser (blinding laser) is working in continuous-wave mode to make the detector enter its linear mode and hence become insensitive to single photons. The second laser is generating optical pulses of $33~\pico\second$ full-width at half-maximum (FWHM) for the tests on detectors D1 and D2 and $161~\pico\second$ for the detectors D3 and D4. The two laser signals are then combined on a $50\!:\!50$ beam splitter.

\section{Detector control}

\subsection{Blinding}
	
First, we test our four devices to see if they are blindable. For that, we increase the continuous-wave optical power $P_\text{blinding}$ arriving on the APD and we measure the rate of detection. Once it reaches 0, the detector is blinded. For our four devices, this happens at an optical power of a few nanowatts and we have tested that they stay blinded up to several milliwatts.
	
\subsection{Forced detections}
\label{sec:control-triggering}

Once Eve has blinded the detector, she can send optical trigger pulses to generate electrical pulses. The amplitude of the signal will be proportional to the energy of the trigger pulse $E_\text{pulse}$. As there is a comparator in the readout circuit, not all pulses are necessarily detected. If the amplitude of the signal is below the comparator threshold, no click will be registered. Therefore, by controlling $E_\text{pulse}$, Eve can force the detector to click with a probability $p \in [0,1]$.  We can then define $E_\text{never}$ as the maximum energy of the optical pulse that never generates a click, and $E_\text{always}$ as the energy above which the detector always clicks. To avoid introducing errors in the key, Eve must carefully choose the energy of her pulse. In the case of the BB84 protocol \cite{bennett1984}, if Eve and Bob measure in different bases, the pulse energy will be divided equally between Bob's two detectors \cite{lydersen2010a}. In this case, Eve does not want Bob's detectors to click, thus she must choose her $E_\text{pulse}<2E_\text{never}$. If Eve's and Bob's bases are the same, all the light will be directed to one detector, which will click with a probability $p$. For short distances, Bob will expect a high detection rate. Eve must then force Bob to click with a high probability, hence the transition region between $E_\text{never}$ and $E_\text{always}$ must be sufficiently narrow. On the other hand, for long-distance QKD, Bob expects a low detection rate, so Eve can afford to have Bob's detector clicking with a low probability.

\floatsetup[figure]{style=plain,subcapbesideposition=top}
\begin{figure}
  \sidesubfloat[]{\includegraphics{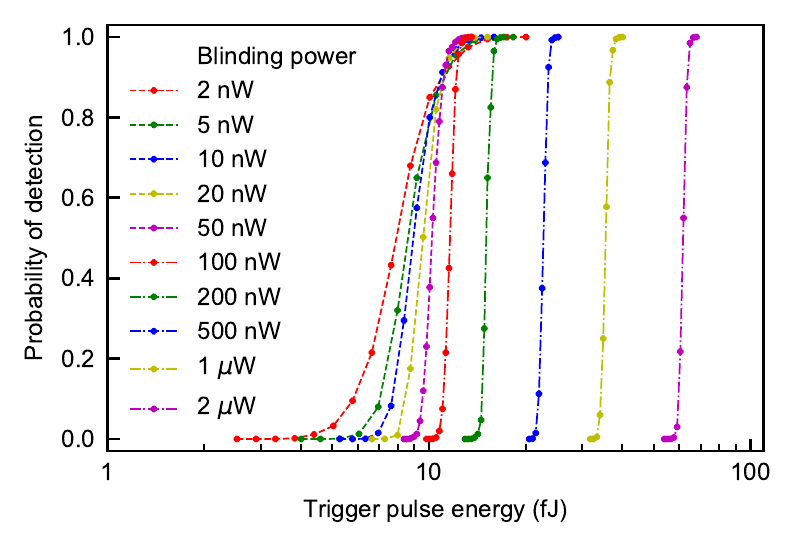}\label{sub:FS-D1}}\\%
  \sidesubfloat[]{\includegraphics{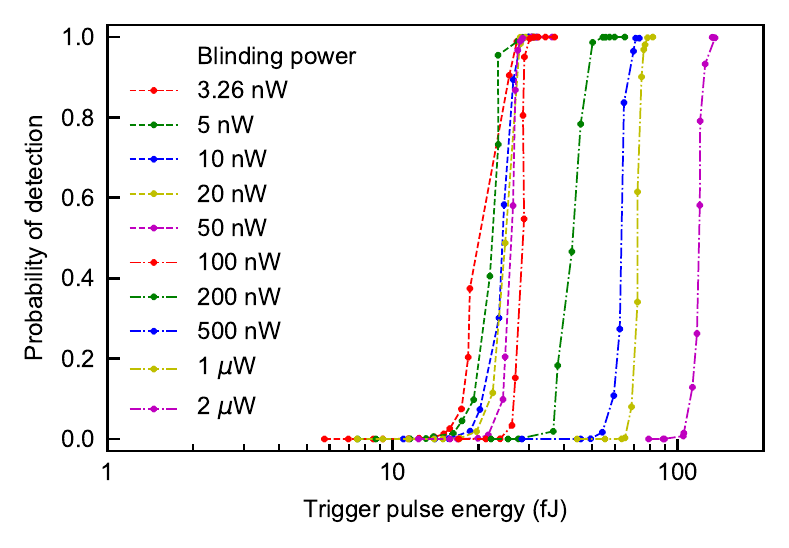}\label{sub:FS-D3}}%
  \caption{Probability to force a detection as a function of the pulse energy for (a)~detector D1 with 10\% photon counting efficiency, and (b)~detector D3 with $2~\volt$ excess bias above $V_\text{br}$. The measurements were made by sending trigger pulses at a frequency of $40~\kilo\hertz$.}
  \label{fig:ProbFS}
\end{figure}

\begin{figure}
  \sidesubfloat[]{\includegraphics{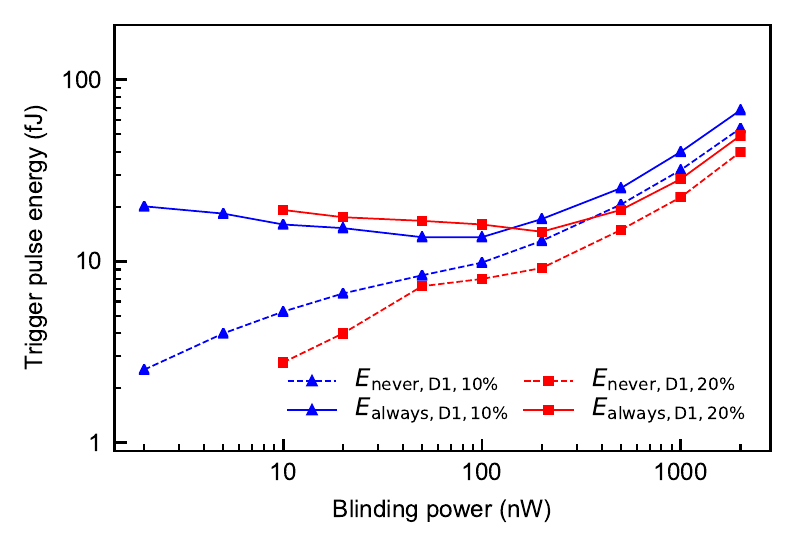}\label{sub:TvB-D1}}\\%
  \sidesubfloat[]{\includegraphics{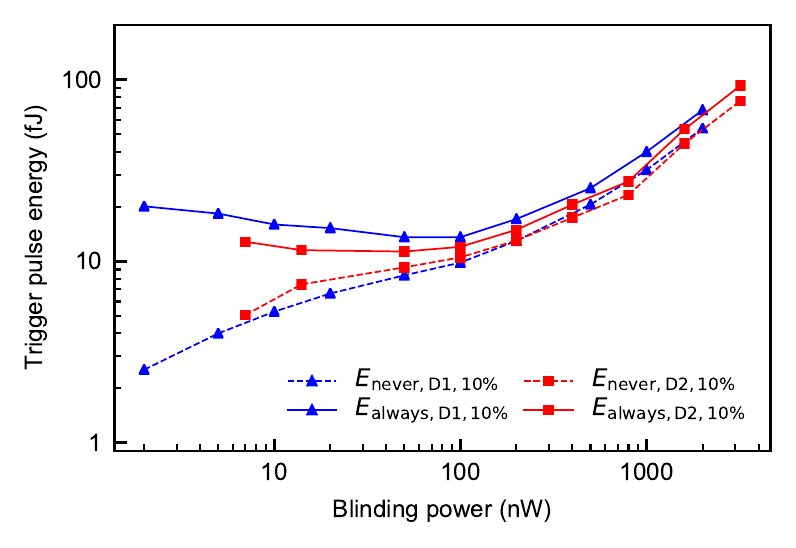}\label{sub:TvB-D1D2}}%
  \caption{Dependence of $E_\text{always}$ and $E_\text{never}$ on the blinding power. (a)~Thresholds for detector D1 with 10\% and 20\% efficiency (corresponding to 1.3~V and 4.1~V excess bias). (b)~Comparison of detectors D1 and D2 with the efficiencies set at 10\%.}
  \label{fig:TrigVsBlinding}
\end{figure}

\begin{figure}
  \sidesubfloat[]{\includegraphics{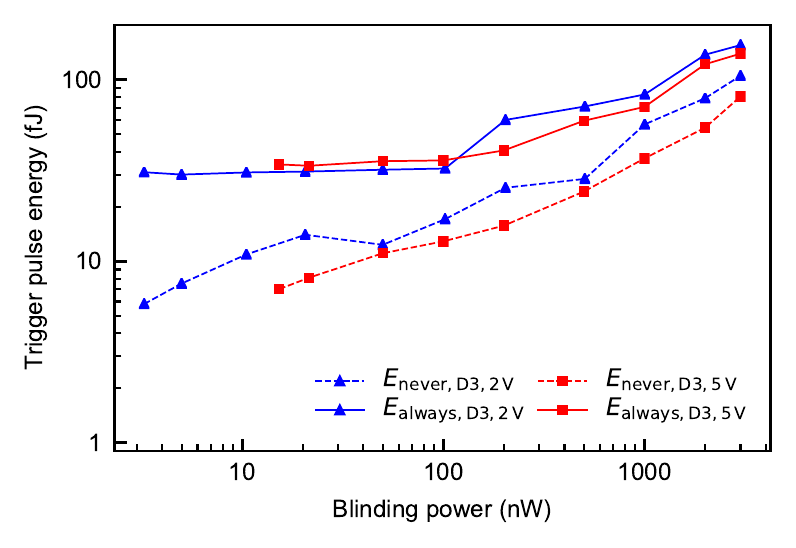}\label{sub:TvB-D3}}\\%
  \sidesubfloat[]{\includegraphics{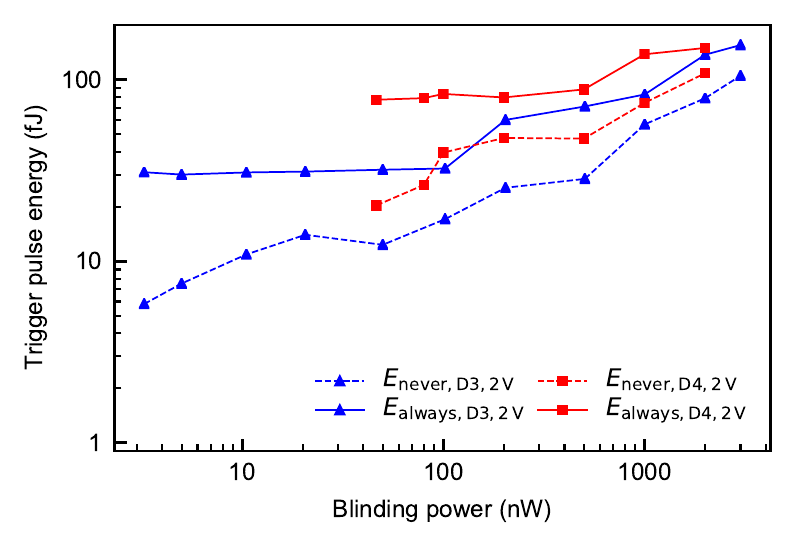}\label{sub:TvB-D3D4}}%
  \caption{Dependence of $E_\text{always}$ and $E_\text{never}$ on the blinding power for the Waterloo detectors. (a)~Thresholds for detector D3 with $2~\volt$ and $5~\volt$ excess bias. (b)~Comparison of detectors D3 and D4 with the same excess voltage of $2~\volt$.}
  \label{fig:D1D2Efficiencies}
\end{figure}

\Cref{fig:ProbFS} shows the probability to get a detection depending on the energy of the trigger pulse for various blinding powers. For this experiment, we set the deadtime $\tau_d$ of the detector at $18~\micro\second$ ($20~\micro\second$), which corresponds to a maximum detection rate of $\sim 55~\kilo\hertz$ ($50~\kilo\hertz$) for detectors D1 and D2 (D3 and D4), and send trigger pulses at a rate of $40~\kilo\hertz$. As we can see in \cref{fig:ProbFS}, there is a transition region where the detection probability monotonically increases from 0 to 1. The changing width of this transition region can be seen in \cref{fig:TrigVsBlinding} for D1 and D2 and in \cref{fig:D1D2Efficiencies} for D3 and D4.

\floatsetup[figure]{style=plain,subcapbesideposition=top}

For high blinding power, the detector is in the linear mode and the APD gain decreases  with the optical power, because the voltage across the APD drops. In order to get the same amplitude of signal at the input of the comparator and get a click, we then need to increase the energy of the trigger pulse. For low blinding power, the detector is in the transition between the linear mode and the Geiger mode \cite{gerhardt2011}. In this region, the probability to generate a macroscopic signal even with a low energy pulse is non-zero, which explains why $E_\text{never}$ decreases when we reduce the blinding power. As seen in \cref{sub:TvB-D1}, when we increase the efficiency of D1 from 10\% to 20\%, the curves are shifted to the right.  This is because the bias voltage is higher for 20\% efficiency hence we need higher $P_\text{blinding}$ to reduced the voltage across the APD to the same value. The detector D3 exhibits a similar effect as seen in \cref{sub:TvB-D3}. Now, if we compare detectors D1 and D2 with the same efficiency, we see in \cref{sub:TvB-D1D2} that both of them have similar triggering energies. The main difference is in the minimum blinding power, which is higher for D2 by a factor of 3. The detectors D3 and D4 require higher triggering energy. This can come from the fact that the detection threshold was set to a higher value due to higher noise in the circuit. We also note that D4 has $\approx 14$ times higher minimum blinding power than D3 (\cref{sub:TvB-D3D4}). Thus for both pairs of detectors, higher minimum blinding power correlates with larger active area. 

\floatsetup[figure]{style=plain,subcapbesideposition=top}

For low blinding power, the transition is too wide for an eavesdropper to attack the entire key in a short distance BB84 protocol \cite{makarov2009}. Eve has then two possibilities: either she increases the blinding power to have a transition region sufficiently narrow, or she attacks only a small part of the key such that Bob's detection rate is not impacted \cite{lydersen2011b}. 

\subsection{Timing jitter}

\begin{figure}
  \sidesubfloat[]{\includegraphics{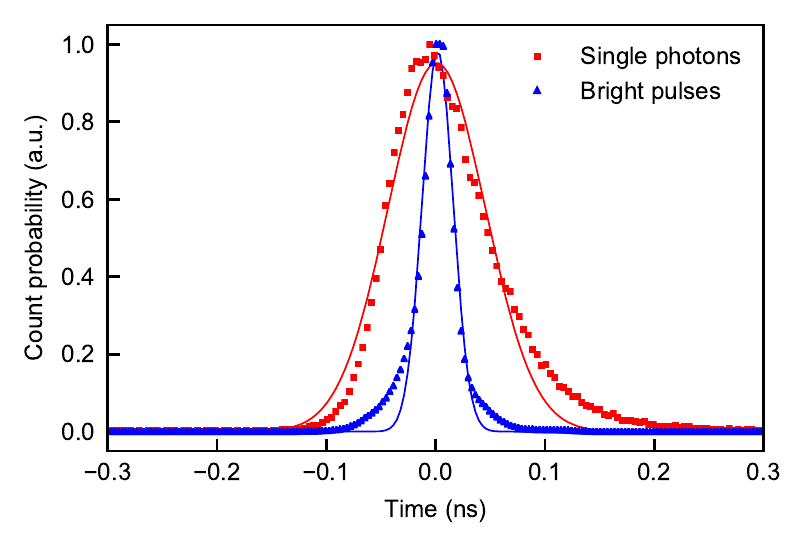}\label{sub:jitter-id220}}\\%
  \sidesubfloat[]{\includegraphics{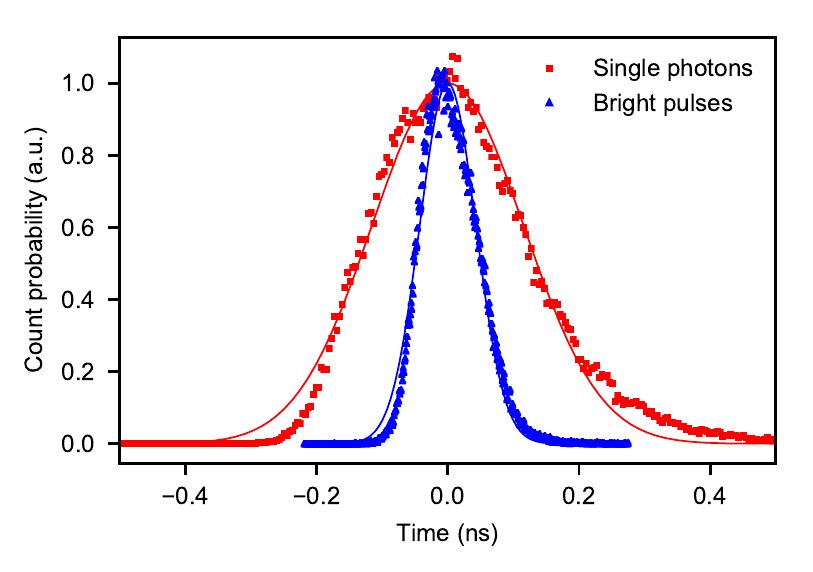}\label{sub:jitter-d3}}%
  \caption{Comparison of the jitter for the detection of a single photon and a bright pulse. The relative time shift between the distributions is not shown; the distributions have been centered. (a)~Jitter of detector D2 with efficiency set at 10\%. The Gaussian fits (solid lines) give a FWHM of $33.4~\pico\second$ for the detection of a faked state ($P_\text{blinding} = 7~\nano\watt$, $E_\text{pulse} = 12.8~\femto\joule$) and $104.9~\pico\second$ for the detection of single photons. (b)~Jitter of detector D3 with $2~\volt$ excess bias. The detection of a faked state ($P_\text{blinding} = 3.3~\nano\watt$, $E_\text{pulse} = 30.9~\femto\joule$) has $100.6~\pico\second$ FWHM and the detection of single photons has $271.8~\pico\second$ FWHM.}
  \label{fig:jitter}
\end{figure}

Another important parameter for Eve is the jitter of the detector's response to her trigger pulse \cite{makarov2009}. Ideally, it should be narrower than single-photon detection jitter. For our measurements, we use a time-correlated single photon counting with the trigger signal for the pulsed laser as a time reference. We perform timing measurements with single photons and bright pulses. For detector D2, we use a $33~\pico\second$ FWHM laser for bright pulses and single-photon jitter measurement; for detector D3, we use $161~\pico\second$ FWHM bright pulses and $147~\pico\second$ FWHM attenuated pulses for single-photon jitter measurement. Results are shown in \cref{fig:jitter}.
 
\floatsetup[figure]{style=plain,subcapbesideposition=top}

As we can see, under control, the jitter of the detection is greatly reduced compared to single-photon detection. Eve is then able to perfectly control in which time bin she wants to make Bob's detector clicks. In order to reproduce the larger jitter of single-photon detections, Eve can artificially increase the jitter of her bright pulses. 

The detector response to the trigger pulse is probably slightly time-shifted relative to its single-photon response. We have not measured this time shift. However this should not hinder Eve in most situations, because she controls the arrival time of her trigger pulse.

\section{Countermeasures}

It is a general assumption in cryptography, called Kerckhoffs’s principle \cite{kerckhoffs1883}, that Eve knows everything about the cryptographic setup and its parameters (detector characteristics under the bright-light control, deadtime, etc.). We therefore have to design a countermeasure that detects the attack even if Eve knows about our countermeasure and tries her best to circumvent it.

One possible way to detect this attack is to monitor the current through the APD. A monitoring circuit is already implemented in ID220. A voltage converter chip biasing the APD has a monitoring pin giving a current equal to 20\% of the average current flowing through the APD, thanks to a current mirror. This current is measured using a 24-bit analog-to-digital converter. In the actual implementation, its value is sampled once per second. We have performed tests of this current monitoring using detector D2 with $\tau_d$ set at $18~\micro\second$. We have first only blinded the detector without sending trigger pulses.

\begin{figure}
	\includegraphics{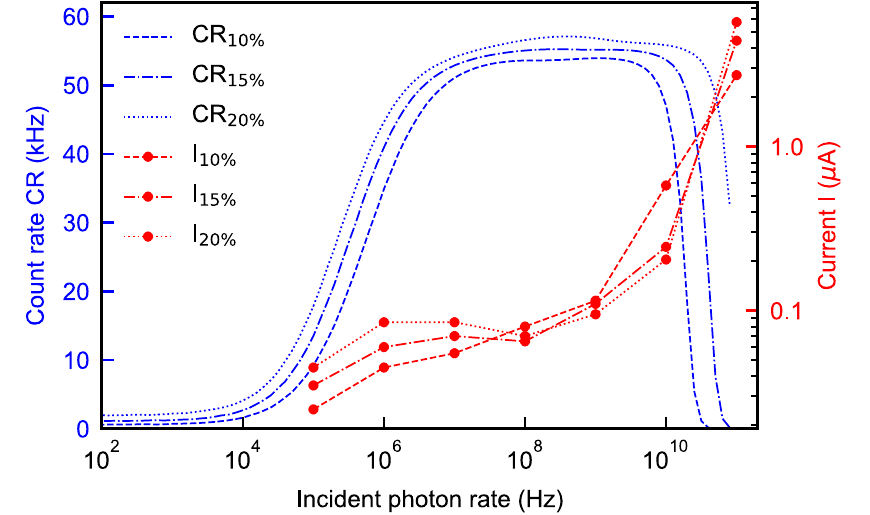}
	\caption{Dependence of the detector D2 count rate and bias current on the incident photon rate. Unlike measurements done with Si detector in \cite{makarov2009}, here we observe a plateau for the count rate due to the deadtime.}
	\label{fig:countrate}
\end{figure}

In normal conditions, the mean current through the APD is very small since the only contribution comes from avalanches due to the detection of a photon. Under control, the blinding laser forces the APD to be continuously conductive. In this case, the mean current should be greater than under normal use. This can be seen in \cref{fig:countrate}. At more than $10^{10}$ incident photons per second, the count rate of the detector drops and reaches 0 (the detector is blinded) while the mean current $I$ increases significantly.  

We have then tested the countermeasure while fully controlling the detector. For that, we used CW blinding and the $33~\pico\second$ FWHM pulsed laser to generate the forced detections. In this case, we see that the mean current through the detector is reduced and depends on the rate of the trigger pulses (see \cref{tab:current-freq}). The explanation comes from the working principle of the detector. Indeed, after a detection, the voltage across the APD is reduced to limit the afterpulsing. During this deadtime ($18~\micro\second$ in our case), the gain of the APD is smaller so the current due to the blinding is reduced. This gives a mean current smaller than that with only the blinding laser.

\begin{table}
\caption{
Current values measured for detector D2 under blinding for different efficiencies and trigger pulse rates.
\label{tab:current-freq}
}
\begin{tabular*}{0.85\columnwidth}[t]{@{\extracolsep{\fill}} c c c}
   \hline
   \hline
   Efficiency (\%) & Pulse rate (kHz) & Current ($\micro\ampere$)\\
   \hline
   10 & 40 & 0.87\\
   10 & 50 & 0.38\\
   10 & 55 & 0.15\\
   20 & 40 & 2.39\\
   20 & 50 & 1.23\\
   20 & 55 & 0.71\\
   \hline
   \hline
\end{tabular*}
\end{table}

The lowest current we could reach was $150~\nano\ampere$ by saturating the detector. This is still higher than the values measured with up to $10^8$ incoming photons per second, which never exceed $100~\nano\ampere$ (\cref{fig:countrate}). By setting the threshold of the current to a proper value (which would depend on $\tau_d$ and detection rate), Bob can thus detect the blinding of his detector by Eve. However this countermeasure is only guaranteed to work provided the blinding is continuous as in our tests, and not a more advanced pulsed one \cite{tanner2014,elezov2019}.

In order to reduce the impact of her attack on the mean photocurrent, Eve has the possibility to take advantage of the detector deadtime to minimize the overall illumination. Indeed, during the deadtime, the voltage across the detector is reduced below $V_\text{br}$ but is still several tens of volts and the blinding laser will unnecessarily generate a current. Hence, by stopping the blinding while the detector is inactive and forcing the detection shortly after its recovery, we can reduce the mean current slightly below $100~\nano\ampere$, making the attack hardly distinguishable from the normal conditions. To detect these short periods of blinding and keep the system secure against the blinding attack, a high-bandwidth measurement is necessary. For that, we use an oscilloscope probe to monitor the output of the bias voltage source (point marked $V_\text{bias}$ in \cref{fig:NFADcircuit}). Due to photocurrent generated by the attack and non-zero output impedance of the bias voltage source, small voltage drops are observed at this point. 

\begin{figure}
	\includegraphics{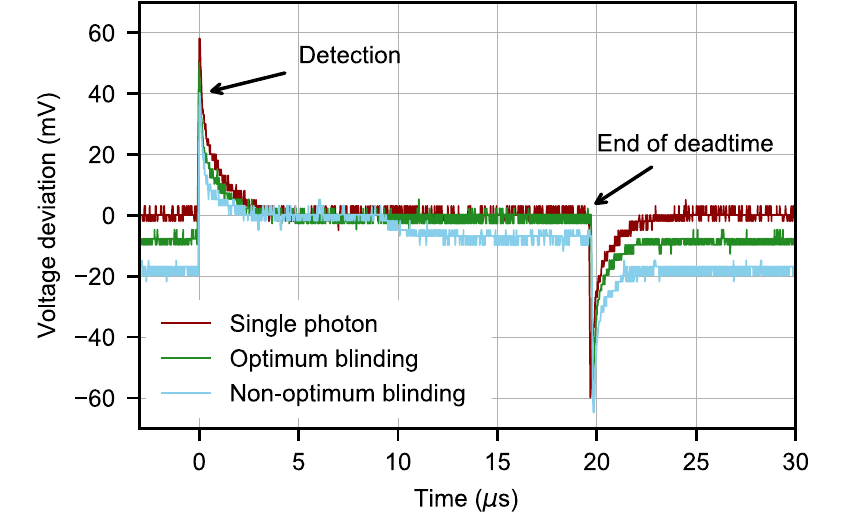}
	\caption{Fluctuations of the bias voltage due to the detection of a single photon (dark red oscilloscope trace) and under the blinding attack (green and blue oscilloscope traces). For an optimum blinding, we use the minimum blinding power and the blinding laser is switched on just at the end of the deadtime. For non-optimum blinding, the laser is switched on in the middle of the deadtime and has higher power.}
	\label{fig:Vdrop}
\end{figure}

\Cref{fig:Vdrop} shows the deviation of $V_\text{bias}$ from its nominal value for detector D2. On each curve, we see two peaks (one positive and one negative) separated by the duration of the deadtime. These are due to high-frequency components of the applied quenching voltage. After the deadtime, we see a voltage drop but only in the case where we blind the detector. This drop comes from the photocurrent induced by the blinding of the detector and lasts as long as the detector is blinded. The deviation of the voltage from its nominal value gives us then information on the state of the detector in real time. The detection of this voltage drop may be used to unveil the presence of an eavesdropper even in the case of more sophisticated attacks such as the one proposed here, and could give Bob information on the bits potentially compromised by this attack.
 
\section{Conclusion}

We have demonstrated the control of four free-running single-photon NFAD detectors by using bright light, which could be used to attack QKD. Mean current monitoring allows us to detect the presence of continuous blinding but might be insufficient in the case of blinding with varying intensity. In the latter case, we have shown that a high-bandwidth measurement of the current flowing through the APD can be used to monitor the state of the detector in real time. This is a step towards constructing a hack-proof single-photon detector for QKD.

\acknowledgments
This project was funded from the European Union's Horizon 2020 programme (Marie Sk\l{}odowska-Curie grant 675662), NSERC of Canada (programs Discovery and CryptoWorks21), CFI, MRIS of Ontario, National Natural Science Foundation of China (grant 61901483), National Key Research and Development Program of China (grant 2019QY0702), and the Ministry of Education and Science of Russia (program NTI center for quantum communications). A.H.\ was supported by China Scholarship Council.


\bibliography{library}

\end{document}